\documentclass[twocolumn,letter]{jpsj3} 
\title{Strong Longitudinal Magnetic Fluctuations \\
near Critical End Point in UCoAl: A $^{59}$Co-NMR Study}

\author{Hiroki \textsc{Nohara}$^{1}$, Hisashi \textsc{Kotegawa}$^{1}$\thanks{E-mail address: kotegawa@crystal.kobe-u.ac.jp}, Hideki \textsc{Tou}$^{1}$, Tatsuma D. \textsc{Matsuda}$^{2}$,\\
 Etsuji \textsc{Yamamoto}$^{2}$, Yoshinori \textsc{Haga}$^{2}$, Zachary \textsc{Fisk}$^{2,3}$, Yoshichika \textsc{\=Onuki}$^{2,4}$,\\
 Dai \textsc{Aoki}$^5$, Jacques \textsc{Flouquet}$^5$}

\inst{$^1$Department of Physics, Kobe University, Kobe 657-8501, Japan\\
$^2$ASRC, JAEA, Tokai, Ibaraki 319-1195, Japan\\
$^3$University of California, Irvine, Calfornia 92697, U.S.A.\\
$^4$Graduate School of Science, Osaka University, Toyonaka, Osaka 560-0043, Japan\\
$^5$INAC/SPSMS, CEA-Grenoble, 17 rue des Martyrs, 38054 Grenoble, France

}

\abst{
We report $^{59}$Co-NMR measurements in UCoAl where a metamagnetism occurs due to enhancement of ferromagnetism by magnetic field.
The metamagnetic transition from a paramagnetic (PM) state to a ferromagnetic state is a first order transition at low temperatures, but it changes to a crossover at high temperatures on crossing the critical end pint (CEP) at $T_{\rm CEP}\sim12$ K.  
The contrasting behavior between the relaxation rates $1/T_1$ and $1/T_2$ suggests that the longitudinal magnetic fluctuation of U moment is strongly enhanced especially near the CEP.
A wide diffusion of the fluctuation from the CEP can be confirmed even in the PM state where the magnetic transition does not occur.

}

\kword{critical end point, UCoAl, metamagnetism, NMR}

\begin{document}
\maketitle

A terminal point of a first order phase transition is called a critical end point (CEP), beyond which the transition changes to a crossover.
Generally a strong first order transition does not provide fluctuations between two phases, but they are expected at the CEP.
The CEP can be seen in various phase diagrams.
One of familiar examples is a liquid-gas transition of water.
The first order phase transition of water encounters the CEP at a high pressure and a high temperature, above which a new exotic state with fluctuations called supercritical water appears.
Another example is a valence transition.
Cerium element shows a first order transition between $\gamma$ phase and $\alpha$ phase under pressure accompanied by a change in the valence.\cite{King,Murani}
The first order line of this $\gamma - \alpha$ transition similarly terminates at a high pressure and a high temperature.
It has been pointed out that a similar valence CEP with the valence fluctuations gives a driving force for exotic superconductivity under pressure in CeCu$_2$Si$_2$.\cite{Holmes,Miyake}
On the other hand, a metamagnetic transition between a paramagnetic (PM) phase and a ferromagnetic (FM) phase occurs when the system is located at the PM side in the vicinity of a first order FM critical point at zero field.\cite{Belitz,Yamada}
The first order metamagnetic transition at low temperatures changes to the crossover at high temperatures on crossing the CEP at $T_{\rm CEP}$.
Interestingly this $T_{\rm CEP}$ decreases under pressure and the FM CEP reaches zero temperature at a high pressure and a high magnetic field, which is generally called quantum critical end point (QCEP).
Such QCEP has been extensively studied theoretically \cite{Belitz,Yamada,Millis,Binz,Yamaji,Imada}, but experimentally the first order plane connecting the CEP at a finite temperature and the QCEP has been drawn schematically because of lack of the systems with large plane.\cite{Pfleiderer_MnSi,Uhlarz}
Recently, sufficiently large first order planes have been experimentally determined in UGe$_2$, UCoAl, and Sr$_3$Ru$_2$O$_7$ ($H \parallel ab$).\cite{Kotegawa,Aoki,Wu}
Because of the benefit of the large plane, we can separate the QCEP from the FM critical point at zero field and address the evolution in the physical quantities over wide pressure and field ranges.
In this letter, before addressing the QCEP, we focus on investigating the magnetic fluctuations near the CEP at the finite temperature and ambient pressure in UCoAl by means of nuclear magnetic resonance (NMR) measurements.

UCoAl crystallizes in the ZrNiAl-type hexagonal structure stacked by U-Co(1) layer and Co(2)-Al layer (see Fig.~2).
UCoAl is a paramagnet but uniaxial pressure effect and chemical substitution have revealed that it is located in the vicinity of a FM critical point.\cite{Mushnikov,Ishi,Andreev2}
It is considered that the FM state at zero field is already suppressed at the negative pressure of $\sim -0.2$ GPa.\cite{Mushnikov}
First order metamagnetic transition to the FM state with $\sim0.3$ $\mu_B/$U occurs at $H_m \sim 0.7$ T for $H \parallel c$ axis at low temperatures.\cite{Andreev}
The FM moment mainly originates in U-$5f$ electrons.\cite{Javorsky}
The transition changes to the crossover above $T_{\rm CEP} \sim 11-13$ K, where the transition occurs at $\sim1$ T.\cite{Mushnikov,Aoki}
Thus UCoAl possesses the CEP at $T_{\rm CEP}\sim11-13$ K and $H_{\rm CEP}\sim1$ T, where the magnetic fluctuations are expected.
NMR measurement has been performed previously to investigate the magnetic fluctuation in UCoAl,\cite{Iwamoto} but measurements near the CEP was not performed sufficiently.
In this letter, we investigated in detail where the fluctuation is strong in the field - temperature phase diagram.
Our NMR results show that the longitudinal magnetic fluctuations strongly develops near the CEP and they widely diffuse even to the PM phase.

We used powdered single crystals with orienting along $H \parallel c$ for NMR measurements.
The single crystal was grown using Czochralski method in a tetra-arc furnace.\cite{Matsuda}
Magnetization ($M$) measurements was performed using a MPMS (Quantum Design).
NMR was performed using a standard spin-echo method.
The spin-spin relaxation time $T_2$ was determined by fitting the spin echo decay to a single exponential function, although the deviation from the single component in the volume of $\sim10$\% was observed near the CEP.

\begin{figure}[htb]
\centering
\includegraphics[width=0.6\linewidth]{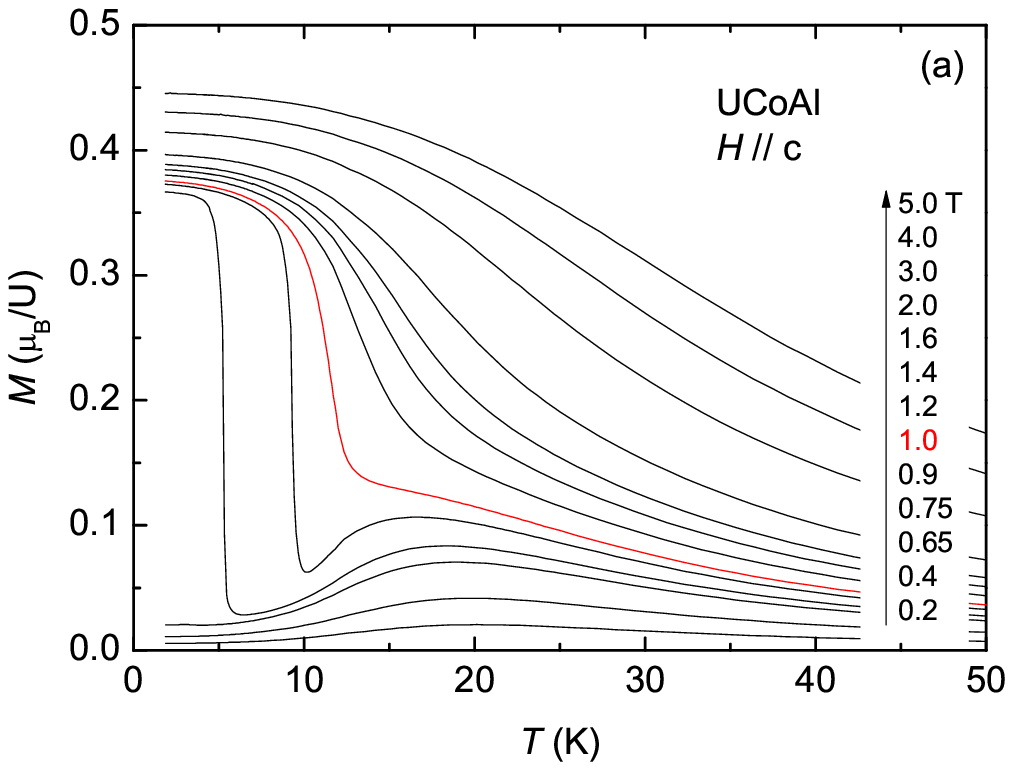}
\includegraphics[width=0.6\linewidth]{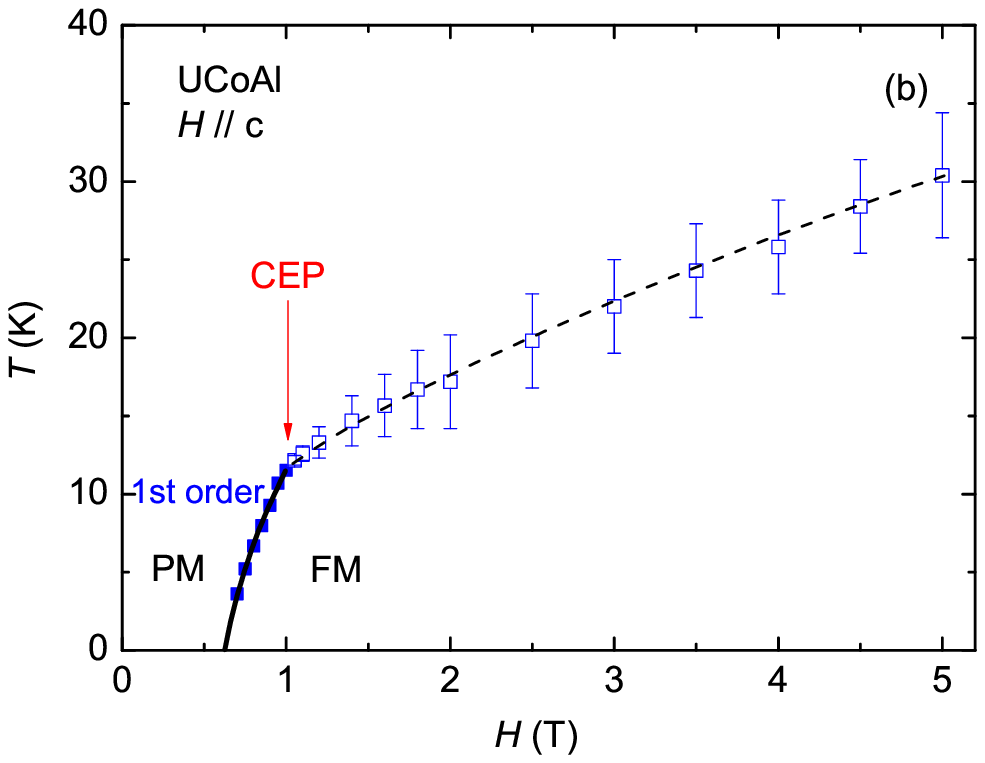}
\caption[]{(color online) (a) Temperature dependence of magnetization at several magnetic fields. (b) Magnetic field - temperature phase diagram for UCoAl. The first order phase transition changes to the crossover at the CEP ($T_{\rm CEP} \sim 12$ K and $H_{\rm CEP} \sim 1.0$ T).
}
\end{figure}

Figure 1(a) shows temperature dependence of $M$ for $H \parallel c$.
Details of magnetization data will be published elsewhere.\cite{Matsuda}
Below 0.65 T, $M$ shows a broad peak around 20 K and the system is in the PM state down to low temperatures.
A discrete increase in $M$ can be seen between 0.75 and 0.9 T due to the first order transition.
The hysteresis in $M(T)$ is observed at these fields and it disappears above 1.0 T.\cite{Matsuda}
The temperature variation of $M$ shows an obvious broadening above 1.0 T changing to the crossover.
We determined a magnetic field - temperature phase diagram for UCoAl as shown in Fig.~1(b).
Both closed and open squares indicate temperature of the peaks in $dM/dT$, where the first order transitions (closed squares) are distinguished by the existence of the hysteresis.
This phase diagram shows that the CEP is located at $T_{\rm CEP} \sim 12$ K and $H_{\rm CEP} \sim 1.0$ T for UCoAl at ambient pressure.

\begin{figure}[htb]
\centering
\includegraphics[width=0.9\linewidth]{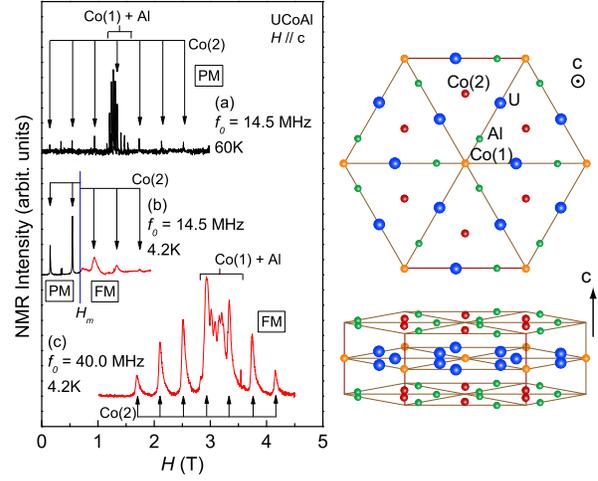}
\caption[]{(color online) (a-c) NMR spectra for UCoAl measured at 60 K and 4.2 K.
The spectrum is discontinuous on crossing $H_m$. The NMR lines of Co(1) site, Al site, and 2 lines of Co(2) site are overlapped at low temperature. (right panel) The crystal structure of UCoAl.
}
\end{figure}

Figures 2(a-c) show the field-sweep NMR spectra measured at different temperatures and resonance frequencies.
UCoAl consists of an U site, two Co sites and an Al site as shown in the right panel of Fig.~2.
At 60 K in the PM state (Fig.~2(a)), the spectrum consists of 19 lines arising from Co(1) and Co(2) sites ($I=7/2$) and Al site ($I=5/2$) as reported previously.\cite{Iwamoto}
Each gyromagnetic ratio $\gamma_N$ is 10.03 MHz/T for $^{59}$Co nuclei and 11.094 MHz/T for $^{27}$Al nuclei.
The sharp peaks ensure the high orientation of the powdered sample along $H \parallel c$.
The satellite lines for Co(2) site are widely separated due to the large quadrupole frequency $\nu_Q$ of about 4.3 MHz compared with other sites.
At low temperature (Fig.~2(b)), we observe two kinds of spectra for the PM state and the FM state because of the discrete enhancement of $M$ by the metamagnetic transition at $H_m \sim 0.7$ T.
When the resonance frequency is increased (Fig.~2(c)), all the lines enter in the FM state, but the lines of Co(1) site, Al site, and 2 lines of Co(2) site are overlapped.
Thus, to avoid the overlap of data, we measured the nuclear spin-lattice relaxation ratio ($1/T_1$) and the nuclear spin-spin relaxation ratio ($1/T_2$) at the satellite line of $-5/2 \leftrightarrow -3/2$ transition of Co(2) site.

\begin{figure}[b]
\centering
\includegraphics[width=0.7\linewidth]{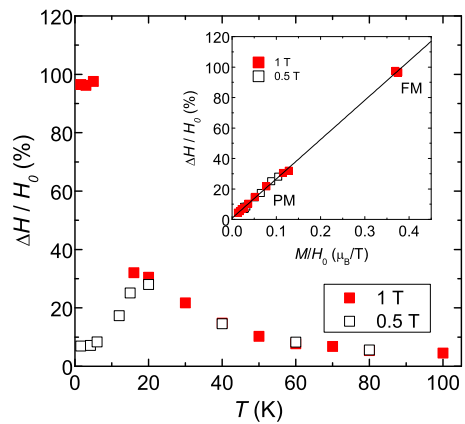}
\caption[]{(color online) Temperature dependence of the NMR shift $\Delta H/ H_0$ at $H_0 = 0.5$ T and 1 T. The inset shows the relationship between the shift and $M/H_0$. The linear relationship gives $A_{spin}^c = 2.58$ (T/$\mu_B$).
}
\end{figure}

Figure 3 shows temperature dependence of the NMR shift ($\Delta H/ H_0$) measured at fixed magnetic fields, where $\Delta H$ is an internal field at Co(2) site and $H_0$ is an applied field.
At 0.5 T, $\Delta H/ H_0$ has a broad maximum at around 20 K as well as temperature dependence of $M$.
The $\Delta H/ H_0$ at 1 T shows a drastic increase between 15 K and 5 K due to the FM transition.
As mentioned later, the small signal by short $T_2$ obstructs to measure the convincing data in this temperature range at 1 T.
The inset shows $\Delta H/ H_0$ vs. $M/H_0$.
In general the shift consists of temperature-dependent spin part, which is proportional to the spin susceptibility, and temperature-independent orbital part.
The linear fitting in the PM range suggests that the orbital part is negligibly small compared with the spin part, giving a hyperfine coupling constant $A_{spin}^c = 2.58$ (T/$\mu_B$).
The data at the FM state also satisfy the linear relationship, indicating the change in the hyperfine coupling constant is small within the experimental precision between the PM state and the FM state, even though the change in the band structure is expected between two phases.

\begin{figure}[htb]
\centering
\includegraphics[width=0.65\linewidth]{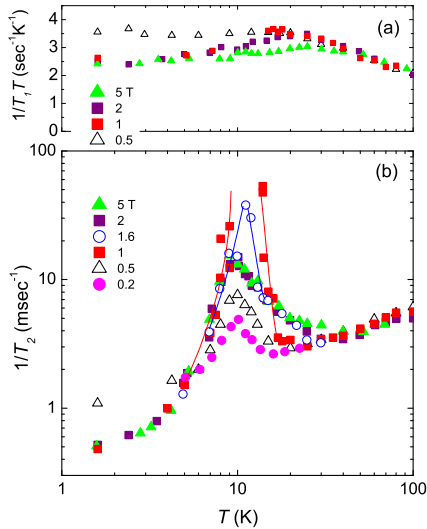}
\caption[]{(color online) Temperature dependence of (a) $1/T_1T$ and (b) $1/T_2$ at several fixed fields. $1/T_2$ has a strong enhancement near the CEP, while $1/T_1T$ does not, suggesting that the observed fluctuations are longitudinal mode of U moments. 
}
\end{figure}

Figure 4 shows temperature dependence of (a) $1/T_1T$, and (b) $1/T_2$ at several fixed fields.
First we should stress that $1/T_2$ shows a strong enhancement in the temperature range of $7-15$ K at $H=H_{\rm CEP}(=1$ T).
We could not measure the relaxation times near 10 K owing to the small signal by short $T_2$ and the broadening of spectrum.
This strong enhancement of $1/T_2$ is drastically suppressed on leaving away from 1 T.
At lower fields (0.2 T and 0.5 T), the peaks in $1/T_2$ are weaken, but still remain at around 10 K even though the magnetic transition does not occur.
The enhancement of $1/T_2$ also survives at higher field  side (1.6 T, 2 T and 5 T) and they are almost field independent above 2 T.
In the previous measurement,\cite{Iwamoto} similar anomaly in $1/T_2$ was observed at 4.5 T and no anomaly was seen at around 10 K at 0 T.
On the other hand the $1/T_1T$ gradually increases with decreasing temperature and shows a constant behavior below 20 K at 0.5 T where susceptibility exhibits the maximum.
$1/T_1T$ is independent of field at high temperatures, but it decreases gradually below $\sim20$ K at 2 T ($\sim30$ K at 5 T), respectively.
The peaks are observed in the wide field range in $1/T_2$, whereas $1/T_1T$ does not show such peaks.
$1/T_1$ and $1/T_2$ are expressed generally as follows.



\begin{eqnarray}
\frac{1}{T_1} = \frac{\gamma_N^2}{2} \int_{-\infty}^{\infty} \langle \delta H^-(t) \delta H^+(0) \rangle \exp(-i \omega_N t) dt, \\
\frac{1}{T_2} = \frac{1}{2T_1} + \frac{\gamma_N^2}{2} \lim_{\omega \rightarrow 0} \int_{-\infty}^{\infty} \langle \delta H^z(t) \delta H^{z}(0) \rangle \exp(-i \omega t) dt.
\end{eqnarray}

Here $\langle \delta H^-(t) \delta H^+(0) \rangle$ [$\langle \delta H^z(t) \delta H^z(0) \rangle$] are time correlation functions for magnetic fluctuations perpendicular [parallel] to the quantum axis of nuclear spin, respectively, and $\omega_N$ is the resonance frequency.
$1/T_1$ corresponds to the magnetic fluctuations perpendicular to the magnetic field, that is, the fluctuations perpendicular to the $c$ axis.
$1/T_2$ indicates the fluctuations along the magnetic field, that is, the fluctuations along the $c$ axis.
Here, as shown in Fig.~2, the position of the Co(2) site is just above the center of triangle formed by three U atoms.
The components along the $c$-axis of U moments induce the internal field only along the $c$ axis at Co(2) site because of the cancelation of the dipole fields.
Thus $1/T_1$ at Co(2) site detects the transverse fluctuation of U moment, while $1/T_2$ at Co(2) site detects the longitudinal fluctuation of U moment.
In the strict sense $1/T_2$ includes the fluctuation of electric field gradient, since it corresponds to the fluctuation of the resonance frequency.
The magnetostriction between the PM phase and the FM phase has been reported.\cite{Aoki}
However, the change in $\nu_Q$ between two phases was less than 0.02 MHz.
This is quite smaller than the change of $\sim6$ MHz in the resonance frequency by the internal field, indicating that the contribution of electron field gradient in $T_2$ is small.
The absence of anomalies in $1/T_1T$ indicates that the observed magnetic fluctuation is the longitudinal fluctuation of U moment along the $c$ axis and the transverse fluctuation of U moment is weak.
This supports that magnetic character near the CEP is a strong Ising-type.

\begin{figure}[htb]
\centering
\includegraphics[width=0.6\linewidth]{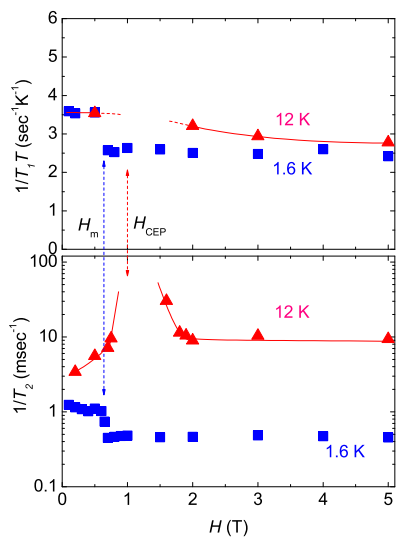}
\caption[]{(color online) Magnetic field dependence of $1/T_1T$ and $1/T_2$. At 12 K $1/T_2$ increases on approaching the CEP. Both $1/T_1T$ and $1/T_2$ show the step-like anomalies at $H_m$ at low temperature.
}
\end{figure}

Figure 5 shows the field dependence of $1/T_1T$ and $1/T_2$ at 1.6 K and 12 K($\sim T_{\rm CEP}$).
At 1.6 K both $1/T_1T$ and $1/T_2$ exhibit step-like anomalies at $H_m$ without the enhancement of fluctuations.
The relationship that $1/T_1T$ is proportional to the square of the density of state at Fermi level $N(E_F)$ gives the reduction of 16\% in $N(E_F)$, which is almost consistent with the specific heat measurements,\cite{Matsuda2,Aoki} as confirmed similarly in the previous study.\cite{Iwamoto}
The $1/T_2$ also shows the drop at $H_m$.
This change might be affected by both the change in the magnitude of fluctuation and the change in the nuclear-dipole interactions because the spectrum in the FM state is broaden.
At 12 K, $1/T_2$ shows the enhancement toward the CEP at 1 T, while it is independent of field above 2 T.
We could not measure $T_1$ in the range of $0.5-2$ T, but the absence of anomaly in $1/T_1T$ as shown in Fig.~4 implies that $1/T_1T$ has no strong enhancement toward the CEP.

\begin{figure}[htb]
\centering
\includegraphics[width=0.7\linewidth]{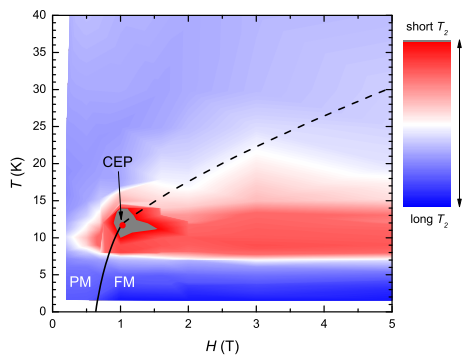}
\caption[]{(color online) The profile of $1/T_2$ in UCoAl. The strong enhancement of $1/T_2$ is observed near the CEP. This longitudinal fluctuation survives in the wide field range. The solid and dotted lines are determined by the peak of $dM/dT$ as shown in Fig.~1(b).
}
\end{figure}

Figure 6 shows the profile of $1/T_2$ in the field - temperature phase diagram for UCoAl.
It is obvious that strong fluctuation is induced near the CEP.
It is suppressed in the first order transition region at low temperatures and the crossover region at high temperatures and high fields.
On the other hand, the fluctuation of the CEP is diffused even to the PM state where $M(T)$ and $\Delta H/H_0$ have no anomaly.
It is an unexpected behavior that field-insensitive fluctuations survive near 10 K up to high magnetic field.
The absence of anomaly in $1/T_1T$ indicates these fluctuations are also the longitudinal mode.
It is worthy to note that $M(T,H)$ has no strong variation in this field and temperature range as shown in Fig.~1(a).
We cannot conclude the origin of these fluctuations, but measurements under pressure will help us to understand the origin.
When looking back on NQR study ($H=0$) under pressure in a similar metamagnetic system UGe$_2$, unknown peak in $1/T_1T$ has been observed in the PM state at 25 K and 1.5 GPa.\cite{Harada}
This is conjectured to be the similar diffused fluctuations from the CEP, since the CEP at 1.5 GPa in UGe$_2$ is located at $\sim20$ K and $\sim0.7$ T.\cite{Valentin}

In summary, we performed $^{59}$Co-NMR measurements in UCoAl with FM CEP at ($H_{\rm CEP} \sim 1$ T, $T_{\rm CEP}\sim12$ K).
The strong enhancement of $1/T_2$ was observed near the CEP.
The absence of anomaly in $1/T_1$ reveals that the fluctuation near the CEP is the longitudinal mode of U moment.
This Ising-type fluctuation is suppressed at the first order regime at low temperatures and the crossover regime at high temperatures, but survives in the wide field range at around 10 K.
The CEP in UCoAl has been reported to attain to zero temperature under pressure.
The evolution of the fluctuation toward the QCEP is a next target.

\section*{Acknowledgement}

We acknowledge H. Harima for valuable discussions.
This work has been partly supported by Grants-in-Aid for Scientific Research (Nos. 19105006, 20740197, 20102005, 22013011, and 22710231) from the Ministry of Education, Culture, Sports, Science, and Technology (MEXT) of Japan.

\end{document}